\documentclass[11pt]{article}
\usepackage{latexsym}  
\usepackage{amssymb}
\usepackage{graphicx}
\usepackage{amsmath}

\begin{document}

\hfill OU-HET 626/2009

\begin{center}
{\Large\bf Yukawaon Model in the Quark Sector}

{\Large\bf and Nearly Tribimaximal Neutrino Mixing}

\vspace{5mm}
{\bf Yoshio Koide}

{\it Department of Physics, Osaka University, 
Toyonaka, Osaka 560-0043, Japan} \\
{\it E-mail address: koide@het.phys.sci.osaka-u.ac.jp}

\date{\today}
\end{center}

\vspace{3mm}
\begin{abstract}
For the purpose of deriving the observed nearly tribimaximal
neutrino mixing, a possible yukawaon model in the quark 
sector is investigated.
Five observable quantities (2 up-quark mass ratios and
3 neutrino mixing parameters $\sin^2 2\theta_{atm}$,
$\tan^2\theta_{solar}$ and $|U_{13}|$) are
excellently fitted by two parameters (one in the up-quark sector 
and another one in the right-handed Majorana neutrino sector).
\end{abstract}

\vspace{2mm}


\vspace{2mm}


\vspace{4mm}

\noindent{\large\bf 1 \ Introduction}

\vspace{2mm}
{\bf 1.1 \ Why do we investigate a quark mass matrix model
for the purpose of discussing the neutrino mixing?}

Recently, on the basis of the so-called ``yukawaon" 
model \cite{mnu-yukawaonPRD08,me-yukawaonPRD09} 
(we will give a short review in the next subsection), 
the author has proposed a curious neutrino mass matrix
form \cite{Koide-O3-PLB08}: although the mass matrix 
$M_\nu$ is given by a conventional seesaw-type form 
$M_\nu = m_D M_R^{-1} m_D^T$,  the Majorana mass 
matrix $M_R$ of the right-handed neutrinos $\nu_R$
is related to a up-quark mass matrix $M_u$ as
$$
M_R \propto M_u^{1/2} M_e + M_e M_u^{1/2} ,
\eqno(1.1)
$$
and the neutrino Dirac mass matrix $m_D$ is given by
$m_D \propto M_e$, where $M_e$ is the charged lepton
mass matrix and $M_u^{1/2}$ is defined by
$\langle M_u^{1/2} \rangle_u ={\rm diag}(\sqrt{m_u},
\sqrt{m_c}, \sqrt{m_t})$ on the diagonal basis of
$M_u$.  [Here and hereafter, we denote a form of a 
matrix $A$ on the diagonal basis of $M_f$ (we refer 
the basis as $f$ basis) as $\langle A\rangle_f$.]
Therefore, the mass matrix $M_\nu$ on the $e$ basis
(the diagonal basis of $M_e$) is given by
$$
\langle M_\nu \rangle_e \propto \langle M_e \rangle_e
\left\{ \langle M_u^{1/2} \rangle_e \langle M_e \rangle_e
+ \langle M_e \rangle_e \langle M_u^{1/2} \rangle_e
\right\}^{-1} \langle M_e \rangle_e .
\eqno(1.2)
$$
In order to calculate the lepton mixing matrix $U$,
we must to know a form of $\langle M_u^{1/2} \rangle_e$.
Since we do not know it at present, 
in Ref.\cite{Koide-O3-PLB08}, on the analogy of a relation
$\langle M_u \rangle_d = V^T(\delta) \langle M_u \rangle_u 
V(\delta)$, where $V(\delta)$ is the Cabibbo-Kobayashi-Maskawa 
(CKM) quark mixing matrix in the standard expression 
\cite{CKM_SD} and the observed value $\delta$ in the 
quark sector is $\delta \simeq 70^\circ$ \cite{PDG08}, 
we have assumed 
$$
\langle M_u^{1/2} \rangle_e = V^T(\pi) 
\langle M_u^{1/2} \rangle_u V(\pi) .
\eqno(1.3)
$$
(Since we assume an O(3) flavor symmetry, the mass 
matrices $M_f$ must be symmetric, so that the 
diagonalization is done by $U_f^T M_f U_f =
M_f^{diag}\equiv \langle M_f \rangle_f$.)
Then, by using the observed up-quark mass ratios 
and CKM mixing parameters, we can obtain the lepton 
mixing matrix $U$.   
Usually, the so-called tribimaximal mixing \cite{tribi} is 
understood based on discrete symmetries, while, in this model, 
the neutrino mass matrix (1.2) can give a nearly tribimaximal
mixing $\sin^2 2\theta_{atm}=0.995$, 
$\tan^2\theta_{solar}=0.553$
and $|U_{13}|=0.001$ \cite{Koide-O3-PLB08}
without assuming any discrete symmetry. 
Note that the matrix $V$ in Eq.(1.3) is not $V(\delta)$ with 
$\delta \simeq 70^\circ$, and it must be $V(\pi)$ in order to 
obtain such the successful results.
This successful results rely on the ad hoc
assumption (1.3).  
We have no theoretical ground for the ansatz (1.3).

However, if we give a quark mass matrix 
model in which quark mass matrices $(M_u, M_d)$ is 
described on the $e$ basis, by using $U_u$ obtained 
from the diagonalization 
$U_u^T \langle M_u \rangle_e U_u = \langle M_u \rangle_u$,
we can calculate the form of $\langle M_u^{1/2}\rangle_e$
as $\langle M_u^{1/2}\rangle_e = U_u \langle M_u^{1/2}
\rangle_u U_u^T$. 
The purpose of the present paper is to investigate 
a possible quark mass matrix model which can give  
reasonable neutrino mixing parameters 
on the basis of a yukawaon model.

In the present paper, we will propose a quark mass 
matrix model which is described on the $e$ basis 
(a diagonal basis of the charged lepton mass matrix): 
$$
\begin{array}{l}
\langle M_u^{1/2} \rangle_e \propto  
\langle M_e^{1/2}\rangle_e \left( {X} + 
a_u   {\bf 1} \right) \langle M_e^{1/2} \rangle_e , \\
\langle M_d \rangle_e \propto \langle M_e^{1/2} \rangle_e 
\left( {X} + a_d  e^{i\alpha_d} {\bf 1} \right) 
\langle M_e^{1/2} \rangle_e , 
\end{array}
\eqno(1.4)
$$
where
$$
X=\frac{1}{3} \left(
\begin{array}{ccc}
1 & 1 & 1 \\
1 & 1 & 1 \\
1 & 1 & 1 
\end{array} \right) , \ \ \ \ 
{\bf 1}= \left(
\begin{array}{ccc}
1 & 0 & 0 \\
0 & 1 & 0 \\
0 & 0 & 1 
\end{array} \right) . 
\eqno(1.5)
$$
(Such a form (1.4) has first been proposed in the context
of the so-called ``democratic seesaw mass matrix model
\cite{democ-seesaw}.)
As we see in the next section, the up-quark mass matrix 
(1.4) can give not only reasonable up-quark mass ratios at 
$a_u \simeq -0.56$, but also reasonable neutrino mixing 
parameters $\sin^2 2\theta_{atm}\simeq 1$, 
$\tan^2 \theta_{solar} \simeq 1/2$ and $|U_{13}| \ll 1$.

\vspace{2mm}
{\bf 1.2 \ What is the yukawaon model? }

Prior to investigating a quark mass matrix model,
let us give a short review of the so-called yukawaon model:
We regard the Yukawa coupling constants in the standard 
model as ``effective" coupling constants $Y_f^{eff}$ 
 ($f=e,\nu,u,d$) in an effective theory, 
and we consider that 
$Y_f^{eff}$ originate in vacuum expectation values (VEVs)
of new gauge singlet scalars $Y_f$, i.e.
$$
Y_f^{eff} =\frac{y_f}{\Lambda} \langle Y_f\rangle ,
\eqno(1.6)
$$
where $\Lambda$ is a scale of the effective theory.
We refer the fields $Y_f$ as 
``yukawaons" \cite{mnu-yukawaonPRD08,me-yukawaonPRD09} 
hereafter.
Note that in the yukawaon model, Higgs scalars are
the same as ones in the conventional model, i.e. we
consider only two Higgs scalars $H_u$ and $H_d$ as an
origin of the masses (not as an origin of the
mass spectra).

The Froggatt-Nielsen model \cite{Froggatt} is well known
as a model which describes masses and mixings by a VEV value
of a scalar $\phi$: The hierarchical structure of the 
masses is explained by a multiplicative structure
$(\langle \phi \rangle/\Lambda)^n$ under a U(1) flavor 
symmetry.
In contrast to the Froggatt-Nielsen model, in the 
yukawaon model, the scalars $Y_f$ have $3\times 3$ 
flavor components, so that they are described, for 
example, as ${\bf 3}\times {\bf 3}^*={\bf 1}+{\bf 8}$ 
of U(3) flavor symmetry or 
$({\bf 3}\times {\bf 3})_S={\bf 1}+{\bf 5}$ of O(3) 
flavor symmetry.
The hierarchical structures of the quark and lepton 
masses are understood from hierarchical eigenvalues
of $\langle Y_f\rangle$, not from the multiplicative
structure $(\langle Y_f\rangle/\Lambda)^n$.
In order to obtain such a hierarchical structure, 
we must build a model with $\det\langle Y_e \rangle/
({\rm Tr[\langle Y_e \rangle]}^3) \ll 1$  (for an example, 
see Ref.\cite{e-Yukawaon09}). 
Hierarchical structures of $\langle Y_f \rangle$
in other sectors will be caused by the hierarchical 
structure of $\langle Y_e \rangle$.

In the present paper, we assume an O(3) flavor symmetry, 
so that would-be Yukawa interactions are given by 
\cite{Koide-O3-PLB08}
$$
H_{Y}= \sum_{i,j} \frac{y_u}{\Lambda} u^c_i(Y_u)_{ij} {q}_{j} H_u  
+\sum_{i,j}\frac{y_d}{\Lambda} d^c_i(Y_d)_{ij} {q}_{j} H_d 
$$
$$
+\sum_{i,j} \frac{y_\nu}{\Lambda} \ell_i(Y_\nu)_{ij} \nu^c_{j} H_u  
+\sum_{i,j}\frac{y_e}{\Lambda} \ell_i(Y_e)_{ij} e^c_j H_d +h.c. 
+ \sum_{i,j}y_R \nu^c_i (Y_R)_{ij} \nu^c_j 
 ,
\eqno(1.7)
$$ 
where $q$ and $\ell$ are SU(2)$_L$ doublet fields, and
$f^c$ ($f=u,d,e,\nu$) are SU(2)$_L$ singlet fields.
All of the yukawaons $Y_f$ ($f=u,d,\nu,e,R$) belong to 
$({\bf 3}\times {\bf 3})_S ={\bf 1}+{\bf 5}$ of O(3). 
In the definition (1.7) of $Y_f$,  we can define 
diagonalization of 
the VEV matrices $\langle Y_f \rangle$ ($f=u,d,e$) as 
$U_f^T \langle Y_f \rangle U_f \propto {\rm diag}
(m_{f1}, m_{f2}, m_{f3})$, 
and that of the seesaw-type neutrino mass matrix 
$M_\nu \propto \langle Y_\nu \rangle^T \langle Y_R 
\rangle^{-1} \langle Y_\nu \rangle$ as
$U_\nu^T  M_\nu  U_\nu = {\rm diag}
(m_{\nu 1}, m_{\nu2}, m_{\nu3})$, 
so that we can express the quark mixing matrix 
$V$ and lepton mixing matrix $U$ as
$V= U_u^\dagger U_d, \ \ {\rm and} \ \ \ U=U_e^\dagger U_\nu$,
respectively.
In the interactions (1.7), in order to distinguish each $Y_f$ 
from others, we have assumed a U(1)$_X$ symmetry in addition 
to the O(3) flavor symmetry, and we have assigned 
U(1)$_X$ charges (``sector" charges, not ``flavor" charges) as 
$Q_X(f^c)=-x_f$, $Q_X(Y_f)= +x_f$ and $Q_X(Y_R)=2x_\nu$.
The SU(2)$_L$ doublet fields $q$, $\ell$, $H_u$ and $H_d$
are assigned as sector charges $Q_X=0$. 
(For the right-handed neutrinos $\nu^c$, we assume 
$Q_X(\nu^c)=Q_X(e^c)$, so that the yukawaon $Y_e$ can 
couple to the Dirac neutrino sector $\ell_i \nu_j^c$ 
as well as $\ell_i e_j^c$, and we can build a model 
without $Y_\nu$ \cite{noYnu}.)

The O(3) flavor symmetry is broken at the energy scale
$\mu=\Lambda$, so that all the yukawaons $Y_f$ have 
VEVs with the order of $\mu=\Lambda$.
(We consider that all components $v_{fi}$ of 
$\langle Y_f \rangle_f ={\rm diag}(v_{f1}, v_{f2}, v_{f3})$
simultaneously have non-zero values at the same energy 
scale $\Lambda$ in spite of the hierarchical structure 
$|v_{f3}|^2\gg |v_{f2}|^2\gg |v_{f1}|^2$.)
A naive estimate of $\Lambda$ leads to $m_\nu \sim 
\langle H_u^0\rangle^2/\Lambda$, so that 
we consider $\mu=\Lambda \sim 10^{15}$ GeV.
Since the O(3) flavor symmetry is completely broken
at $\mu=\Lambda$, the effective coupling constants 
$Y_f^{eff}=(y_f/\Lambda)\langle Y_f \rangle$ evolve as 
in the standard model below the scale $\Lambda$.

In a supersymmetric (SUSY) yukawaon  model, 
VEV structures of the yukawaons are obtained from 
SUSY vacuum conditions for a superpotential $W$. 
As a result, a VEV structure of $\langle Y_f \rangle$ 
in an $f$ sector is described in terms of other yukawaon 
VEV matrices. 
In other words, a VEV structure of $\langle Y_f \rangle$
is given by observed mass matrices in other sectors. 
(The relation (1.1) is one of the examples.) 
This is just a characteristic feature in the yukawaon
approach.

For example, in the present scenario, we have assumed
superpotential terms
$$
W_e = \mu_e [Y_e \Theta_e] +
\lambda_e [\Phi_e \Phi_e \Theta_e]  , 
\eqno(1.8)
$$
and, from the SUSY vacuum coditions 
$\partial W/\partial \Theta_e=0$, we obtain a bilinear 
mass relation
$$
\langle Y_e\rangle = -\frac{\lambda_e}{\mu_e}
\langle \Phi_e\rangle \langle \Phi_e\rangle .
\eqno(1.9)
$$
Thus, the mass matrix $M_e^{1/2}$ in Eq.(1.2) is 
interpreted as $\langle \Phi_e\rangle$ in the yukawaon
model.
We call $\Phi_e$ as ``ur-yukawaon", because the VEV of
$\Phi_e$ plays a role in giving VEV spectrum of $Y_e$.
Here and hereafter, for convenience, we denote 
Tr$[A]$ as $[A]$ simply.
(We can show $\langle\Theta_e\rangle =0$ by solving
SUSY vacuum conditions for all fields simultaneously.
Hereafter, we will use a character $\Theta$ for a 
field whose VEV is $\langle \Theta\rangle =0$.)
Also, we consider the existence of an ``ur-yukawaon"
$\Phi_u$ which has a relation 
$\langle Y_u \rangle \propto \langle \Phi_u\rangle
\langle \Phi_u\rangle$ similar to (1.8).

The relation (1.1) is naively  
translated to a relation 
$\langle Y_R \rangle \propto \langle \Phi_u\rangle
\langle Y_e \rangle + \langle Y_e \rangle \langle 
\Phi_u\rangle$ in the yukawaon model, but, as stated
in the next section, the relation (1.1) will be slightly 
modified (a term with same U(1)$_X$ charge will be added 
with an additional parameter $\xi$).
As a result, as we see in Sec.2, we can excellently fit 5 
observable quantities (2 up-quark mass ratios and 3 neutrino 
mixing parameters) by the two parameters $a_u$ and $\xi$.
Considering such the phenomenological success of the model, 
in Sec.3, a possible yukawaon model in the quark sector 
and neutrino sector is discussed. 
Finally, Sec.4 is devoted to a summary and concluding 
remarks.

In the present model, 2 down-quark mass ratios 
and 4 CKM mixing parameters are described by the remaining 
2 parameters $a_d$ and $\alpha_d$ after $a_u$ was fixed by
the observed up-quark mass ratios.
As seen in Sec.2, the results are still unsatisfactory.
We need a further improvement for down-quark sector.
In the present paper, we do not discuss any improvement 
in the down-quark sector.

\vspace{3mm}

\noindent{\large\bf 2 \ Phenomenological study}

In the present section, we give numerical studies
based on phenomenological mass matrices (1.2) and (1.4).  
Since the aim of the present paper is not to obtain 
the best fit values of the parameters in the model, 
and it is to see a main framework of the model qualitatively, 
an energy scale used in the present evaluations
is not required to be so rigorous. 
Exactly speaking, VEV relations in the present
model are valid at $\mu=\Lambda \sim 10^{15}$ GeV. 
Since the O(3) flavor symmetry is completely broken
at $\mu=\Lambda$, the effective coupling constants 
$Y_f^{eff}=(y_f/\Lambda)\langle Y_f \rangle$ evolve as 
in the standard model below the scale $\Lambda$.
We know that the quark mass ratios are
not so sensitive to the energy scale \cite{q-mass}.
As seen in (2.1) and (2.6) below, the observed quark
mass values \cite{q-mass} still have large errors
and, besides, they are highly dependent on the 
value of $\tan\beta$ in the SUSY model.
Since our concern is in the quark mixing matrix 
$V=U_u^\dagger U_d$ and lepton mixing matrix 
$U=U_e^\dagger U_\nu$,
for simplicity, we evaluate the mass matrices (1.4) 
at the energy scale $\mu=m_Z$,  
so that numerical results in the present section 
should not be taken too strictly.
In the present model, quark mass matrices are 
described in terms of the charged lepton masses, 
so that, for numerical estimates, we will also use values 
of the running masses $m_e(\mu)$, $m_\mu(\mu)$ and 
$m_\tau(\mu)$ at $\mu=m_Z$.
Although the yukawaon model \cite{me-yukawaonPRD09}
was first proposed with the aim of understanding 
the well-known charged lepton mass relation \cite{Koidemass}, 
in the present paper, we do not adopt such a yukawaon 
model in the charged lepton sector. 
We will use the observed values \cite{q-mass} of 
the running charged lepton masses as input values,
which do not satisfy the charged lepton mass relation.

First, we search for a value of the parameter
$a_u$ which can give reasonable predicted values for 
the up-quark mass ratios \cite{q-mass} 
${m_u}/{m_c} = 0.0021^{+0.0013}_{-0.0008}$ 
($0.0021^{+0.0012}_{-0.0009}$) and 
${m_c}/{m_t} = 0.0036^{+0.0006}_{-0.0005}$ 
($0.0026^{+0.0007}_{-0.0006}$)
at $\mu=m_Z$ (at $\mu=\Lambda_{GUT}=2\times 10^{16}$ GeV
with $\tan\beta=10$).
Although the ratio $m_u/m_t$ is affected by
renormalization group equation (RGE) effects, 
the effects are not so essential in this rough estimations
in the present paper.
Therefore, we will use values at $\mu=m_Z$:
$$ 
\sqrt{m_u/m_c}=0.045^{+0.013}_{-0.010}, \ \ \  
\sqrt{m_c/m_t}=0.060\pm 0.005 .
\eqno(2.1)
$$
We find that predicted values of the up-quark
mass ratios at $a_u \simeq -0.56$ is in favor of 
the observed up-quark mass ratios (2.1):
$v_{u1}/v_{u2}=(-0.0355,-0.0425,-0.0514)$ 
and $v_{u2}/v_{u3}=(-0.0654,-0.0570,-0.0495)$
for $a_u=(-0.55,-0.56,-0.57)$. 
However, if we naively regard $\langle \Phi_u \rangle_e$ 
given in Eq.(1.4) as $\langle M_u^{1/2} \rangle_e$ in 
Eq.(1.2), we cannot obtain favorable values of the 
neutrino mixing parameters for any values of $a_u$.,

Note that signs of the eigenvalues $(v_{u1}, v_{u2}, v_{u3})$
of $\langle\Phi_u\rangle$ are $(+,-,+)$ for the parameter value 
$a_u \sim -0.56$, while, in Ref.\cite{Koide-O3-PLB08}, 
we have used positive values for 
 $\langle M_u^{1/2} \rangle_e={\rm diag}(\sqrt{m_u}, 
\sqrt{m_c}, \sqrt{m_t})$.
Therefore, by introducing an O(3) ${\bf 1}+{\bf 5}$ field
$P_u$ whose VEV is given by
$$
\langle P_u \rangle_u = \mu_p {\rm diag}(+1,-1,+1) ,
\eqno(2.2)
$$
we can express $\langle M_u^{1/2}\rangle$ as 
$\langle M_u^{1/2}\rangle_u \propto \Phi_u\rangle_u 
\langle P_u\rangle_u$, so that the relation (1.2) is given by
$$
\langle Y_R\rangle_e \propto \langle Y_e\rangle_e
\langle P_u\rangle_e \langle \Phi_u\rangle_e
+\langle \Phi_u\rangle_e \langle P_u\rangle_e
\langle Y_e\rangle_e ,
\eqno(2.3)
$$
where $\langle \Phi_u\rangle_e$ and $\langle P_u\rangle_e$
are given by 
$$ 
\langle \Phi_u\rangle_e=U_u \langle \Phi_u \rangle_u U_u^T,
\ \ \ 
\langle P_u\rangle_e= U_u \langle P_u\rangle_u U_u^T ,
\eqno(2.4)
$$
respectively.
However, when we assign U(1)$_X$ charges which satisfy
$Q_X(Y_R)= Q_X(Y_e)+Q_X(\Phi_u)+Q_X(P_u)$, 
we must also take terms $P_u Y_e \Phi_u + \Phi_u Y_e P_u$ 
into consideration as well as the terms 
$Y_e P_u \Phi_u +\Phi_u P_u Y_e$, because
they have the same U(1)$_X$ charges.
Thus, the relation (2.3) must be modified as
$$
Y_R \propto Y_e P_u \Phi_u + \Phi_u P_u Y_e 
+\xi (P_u Y_e \Phi_u + \Phi_u Y_e P_u) .
\eqno(2.5)
$$

We find that predicted value of $\tan^2 \theta_{solar}$
is highly dependent on the parameter $\xi$, although 
the values $\sin^2 2\theta_{atm}$ and $|U_{13}|$ are 
not sensitive to $\xi$ as far as $|\xi|$ is small.
We show $\xi$-dependence of the neutrino mixing 
parameters in Table 1.
As seen in Table 1, the cases $\xi=+0.0005$ and 
$\xi=-0.0012$ can excellently fit the observed values
 $\sin^2 2\theta_{atm}=1.00_{-0.13}$ \cite{MINOS}
and $\tan^2 \theta_{solar}=0.469^{+0.047}_{-0.041}$
 \cite{SNO08}. 
 At present, the reason why the parameter value of $\xi$ 
is so small is unknown.
This may be explain by a hidden symmetry which is unbroken
in the terms without $\xi$, but is broken in those with
$\xi$.
This is an open question at present.
 
\begin{table}
\caption{$\xi$-dependence of neutrino mixing parameters.
The parameter $a_u$ is fixed at $a_u \sim -0.56$. }

\vspace{2mm}
\begin{tabular}{|c|ccc|} \hline
$\xi$ & $\sin^2 2\theta_{atm}$ & 
$\tan^2 \theta_{solar}$ & $|U_{13}|$ \\ \hline
0     & 0.9848 & 0.7033 & 0.0128 \\ \hline
$+0.004$  &  0.9825 & 0.4891 & 0.0123 \\
$+0.005$ &  0.9819 & 0.4486 & 0.0122 \\ 
$+0.006$ &  0.9812 & 0.4123 & 0.0120 \\ \hline
$-0.0011$ & 0.9897 & 0.4854 & 0.0142 \\
$-0.0012$ & 0.9900 & 0.4408 & 0.0143 \\
$-0.0013$ & 0.9904 & 0.4008 & 0.0144 \\
\hline
\end{tabular} 
 \end{table}

Next, we calculate down-quark mass ratios and 
CKM matrix parameters for the model (1.4).
The observed running down-quark mass ratios \cite{q-mass} 
are
$$
\frac{m_d}{m_s} =
\begin{array}{l}
 0.053^{+0.051}_{-0.029}\\
(0.054^{+0.058}_{-0.030})
\end{array} , \ \ \ 
\frac{m_s}{m_b} = 
\begin{array}{l}
0.019 \pm0.006 \\
(0.017^{+0.006}_{-0.005} )
\end{array} , 
\eqno(2.6)
$$
at $\mu=m_Z$ (at $\mu=\Lambda_{GUT}=2\times 10^{16}$ GeV
with $\tan\beta=10$) and the observed CKM mixing parameters
\cite{PDG08} are $|V_{us}|=0.2255 \pm 0.0019$, 
$|V_{cb}|=0.0412\pm 0.0011$, $|V_{ub}|=0.00393 \pm 0.00036$
and $|V_{td}|=0.0081 \pm 0.025$.
In Table 2, we demonstrate predicted values of the
CKM mixing parameters versus $(a_d, \alpha_d)$.
Here, we have taken $a_u=-0.56$ which can give 
reasonable up-quark mass ratios.

As seen in Table 2, the case with $a_u=-0.56$ and 
$(a_d, \alpha_d)=(-0.63, 8^\circ)$ can roughly give
reasonable values of the down-quark mass ratios and  
$|V_{us}|$, but $|V_{cb}|$, $|V_{ub}|$ and $|V_{td}|$
are considerably larger than the observed values.
For reference, we list a case of $a_u=-0.58$ and 
$(a_d, \alpha_d)=(-0.63, 2^\circ)$ in Table 3.
The case can give reasonable values of $|V_{us}|$
and $|V_{cb}|$, but the predicted quark mass ratios
are in poor agreement with experiments. 
The present model will need a further improvement,
as far as the down-quark sector is concerned.
However, from the qualitative point of view, it worthwhile
noticing  that 
the model can roughly predict not only the quark mass ratios 
but also the CKM mixing parameters and neutrino oscillation 
parameters by using only the 4 input parameters $a_u$, $\xi$
and $a_d e^{i \alpha_d}$.

\begin{table}
\caption{CKM mixing parameters versus $(a_d, \alpha_d)$}.

\vspace{2mm}
\begin{tabular}{|c|cc|cc|cccc|} \hline
$a_u$ & $a_d$ & $\alpha_d$ & $|m_{d1}/m_{d2}|$ & 
$|m_{d2}/m_{d3}|$ & $|V_{us}|$ & $|V_{cb}|$ & 
 $|V_{ub}|$ & $|V_{td}|$ 
\\ \hline
$-0.56$ & $-0.620$  &$4^\circ$ & 0.1078 & 0.0273 & 
0.2035 & 0.0666 & 0.0101 & 0.0178 \\
$-0.56$ & $-0.625$  &$6^\circ$ & 0.0783 & 0.0313 & 
0.2187 & 0.0818 & 0.0123 & 0.0190 \\
$-0.56$ & $-0.630$  &$8^\circ$ & 0.0542 & 0.0362 & 
0.2222 & 0.0977 & 0.0146 & 0.0194 \\
$-0.58$ & $-0.630$  &$2^\circ$ & 0.1959 & 0.0195 & 
0.2272 & 0.0448 & 0.0088 & 0.0163 \\
\hline
\end{tabular}
\end{table}


\vspace{3mm}

\noindent{\large\bf 3 \ Yukawaons in the quark sector}

In the previous section, we have obtained successful 
predictions on the basis of the phenomenological
mass matrices (1.4) and (2.5).
For the phenomenological mass matrix (2.5), we assume 
the following superpotential terms 
$$
W_R = \mu_R [Y_R \Theta_R] + \frac{\lambda_R}{\Lambda}
\left\{ [(Y_e P_u \Phi_u+\Phi_u P_u Y_e) \Theta_R] 
+\xi [(P_u Y_e \Phi_u +\Phi_u Y_e P_u) \Theta_R] \right\}.
\eqno(3.1)
$$
Here, since the VEV matrix $\langle P_u\rangle$ is diagonal
in the $u$ basis (a diagonal basis of $\langle Y_u\rangle$)
as given in Eq.(2.2), the VEV matrix $\langle P_u\rangle$ 
must be commutable with $\langle \Phi_u\rangle$.
Moreover, in order to give the form (2.2),  since 
$[\langle P_u\rangle]=\mu_p$,
$\frac{1}{2}([\langle P_u\rangle]^2-[\langle P_u\rangle
\langle P_u\rangle])=-\mu_p^2$ and
$\det \langle P_u\rangle =-\mu_p^3$, 
the VEV matrix $\langle P_u\rangle$ must satisfy a cubic 
equation
$$
\langle P_u\rangle^3 -\mu_p \langle P_u\rangle^2 
-\mu_p^2 \langle P_u\rangle +\mu_p^3 {\bf 1}=0 .
\eqno(3.2)
$$
Therefore, we assume superpotential terms for $P_u$
$$
W_P= \lambda_P [(\Phi_u P_u - P_u \Phi_u) \Theta_P]
+ \varepsilon_{SB} \frac{\lambda'_P}{\Lambda}[(P_u P_u P_u 
-\mu_p P_u P_u -\mu_p^2 P_u +\mu_p^3 )\Theta'_P] .
\eqno(3.3)
$$
Note that the second term in (3.3) cannot conserve the 
U(1)$_X$ charges.
(The superpotential terms (3.3) can conserve the $R$ 
symmetry, because the fields $\Theta_P$ and $\Theta'_P$
are assigned as $R$ charges $R=2$, while $\Phi_u$ and 
$P_u$ are assigned as $R=0$.)
We assume that such U(1)$_X$ symmetry breaking terms are
suppressed by a small factor $\varepsilon_{SB}$. 

On the other hand, superpotential in the quark sector
is given as follows:
$$
W_q = \mu_u^X [\Phi_u\Theta_u^X] +\mu_d^X [Y_d \Theta_d^X]
+ \sum_{q=u,d} \frac{\xi_q}{\Lambda} [\Phi_e 
(\Phi_{Xq} + a_q E_q ) \Phi_e \Theta_q^X] ,
\eqno(3.4)
$$
where the parameter $a_u$ is real, but $a_d$ is complex.
Here the VEV matrices $\langle \Phi_{Xq} \rangle_e$ and
$\langle E_q \rangle_e$ are given by
$$
\langle \Phi_{Xq} \rangle_e \propto X, \ \ \
\langle E_q \rangle_e \propto {\bf 1} ,
\eqno(3.5)
$$
and the factor $(\Phi_{Xq} + a_q E_q )$ plays a role in 
breaking the O(3) flavor symmetry into a discrete symmetry
S$_3$ at the $e$ basis. 
If we regard $\Phi_{Xq}$ and $E_q$ as 
$\Phi_{Xu}=\Phi_{Xd}\equiv\Phi_X$
and $E_u=E_d\equiv E$, we will obtain an unwelcome relation 
$Q_X(\Phi_u)=Q_X(Y_d)$. 
Therefore, we must distinguish  $(\Phi_{Xu}, E_u)$ 
from $(\Phi_{Xd}, E_d)$. 
However, if we replace $\mu_u^X$ and $\mu_d^X$ in (3.4) 
with $[Y_d]$ and $[\Phi_u]$, respectively, we can regard 
$\Phi_{Xq}$ and $E_q$ as $\Phi_{Xu}=\Phi_{Xd}\equiv\Phi_X$
and $E_u=E_d\equiv E$ without considering two sets 
$(\Phi_{Xu}, E_u)$ and $(\Phi_{Xd}, E_d)$.
Superpotential terms for the field $\Phi_X$ are 
given by
$$
W_X=
\varepsilon_{SB} \left\{ \frac{\xi_X}{\Lambda} 
[\Phi_X \Phi_X \Phi_X \Theta_X] +
\lambda_X [\Phi_X \Phi_X \Theta_X] \right\},
\eqno(3.6)
$$
with $(\lambda_X/\xi_X) \Lambda =- [\Phi_X]$,
because 
$$
[X]= 1, \ \ \frac{1}{2} \left( [X]^2 -[XX] \right) =0 , 
\ \ \ \det X =0.
\eqno(3.7)
$$
Also, superpotential terms for the field
$E$ are given by
$$
W_E =
\varepsilon_{SB} \left( \mu_E [E \Theta_E] + \mu_E^2 [\Theta_E] 
\right).
\eqno(3.8)
$$
Note that we cannot write the superpotential terms (3.6) 
and (3.8) without breaking the U(1)$_X$ symmetry
explicitly, so that we have added a factor $\varepsilon_{SB}$
in the superpotential terms (3.6) and (3.8).

In the superpotential terms (3.3), (3.6) and (3.8), we 
have assumed some specific forms of U(1)$_X$ symmetry
breaking terms.
As such symmetry breaking terms, in general, not only
those given in (3.3), (3.6) and (3.8) but also many
other terms are allowed.
Therefore, at present, the forms of the $\varepsilon_{SB}$-terms 
are only ones which are required from a phenomenological 
point of view.

\vspace{3mm}

\noindent{\large\bf 4 \ Concluding remarks}

In conclusion, for the purpose of explaining the observed
tribimaximal neutrino mixing, we have proposed a yukawaon 
model (1.4) [(3.4)] in the
quark sector, where the O(3) symmetry is broken into 
S$_3$ by the VEV $\langle \Phi_X \rangle_e$ on the 
$e$ basis (the diagonal basis of $\langle Y_e \rangle$).
The up-quark mass matrix given in Eq.(1.4) includes one
parameter $a_u$.
In the seesaw-type neutrino mass matrix 
$M_\nu = m_D M_R^{-1} m_D$, the Majorana mass matrix
of the right-handed neutrinos $M_R \propto \langle Y_R \rangle$
is related to the up-quark mass matrix 
$M_u \propto \langle Y_u \rangle \propto 
\langle \Phi_u \rangle \langle \Phi_u \rangle$
as given in Eq.(2.5) which includes one parameter $\xi$.
The parameters $a_u$ and $\xi$ can describe 5 observables 
(2 up-quark mass ratios and 3 neutrino oscillation parameters).
The numerical results for the neutrino oscillation 
parameters excellently gives nearly tribimaximal mixing 
as shown in Table 1.

So far, we did not mention neutrino mass ratios, i.e.
$R=\Delta m^2_{solar}/\Delta m^2_{atm}$.
The ratio $R$ can favorably be fitted by adding 
\cite{mnu-yukawaonPRD08} a term 
$\lambda'_R [Y_\nu Y_\nu \Theta_R]$ to the superpotential 
terms $W_R$ (3.7) [(3.1)], or by adding 
\cite{noYnu} a term $(y'_R/\Lambda)\nu^c Y_\nu Y_\nu \nu^c$
to the would-be Yukawa interaction (1.7).
Those terms play a role in shifting neutrino masses commonly 
without affecting neutrino mixing.
However, the added parameter is exactly fixed by the observed
value of $R$, so that there is no prediction in the neutrino
sector. 

Although the present model with (1.4) and (2.5) can describe   
values of 11 observables (4 quark mass ratios, 4 CKM 
mixing parameters, and 3 neutrino oscillation 
parameters) by adjusting 4 parameters (1 and 2 in the 
up- and down-quark sectors, respectively, and 1 in the 
$Y_R$ sector), the fitting for down-quark masses and 
the CKM mixing parameters is poor as seen in Table 2, 
although the results are qualitatively not so bad.
The two parameter description in the down-quark sector 
is too tight. 
We think that the present model is a step in the
right direction to a unified yukawaon model. 
In the next step, we will investigate a further improvement
of the down-quark sector.

In the present model, we have assumed an O(3) flavor symmetry.
The VEV relations which are derived from the superpotential
under the O(3) symmetry are valid only in specific flavor
bases which are connected by orthogonal transformations. 
We have regarded the $e$-basis (the diagonal basis of
$\langle Y_e\rangle$) as a specific basis, in which 
the relations from the O(3) symmetry are valid and the
VEV matrix $\langle \Phi_X\rangle$ takes a simple form (3.5). 

The present approach based on a yukawaon model  
seems to provide a new view to a unified description
of the masses and mixings  differently from conventional
mass matrix models.
Under few parameters in the quark sector, the model can predict 
not only all quark and lepton mass ratios
but also the CKM matrix parameters and neutrino 
oscillation parameters, although an improvement is 
still needed as far as the down-quark sector is concerned.  
It is worthwhile taking the present model seriously
as a promising model which can give a unified 
description of the quark and lepton masses and
mixings.
Further development of the model is expected.

\vspace{6mm}

\centerline{\large\bf Acknowledgments}

The author would like to thank T.~Yamashita, 
M.~Tanaka, N.~Uekusa and H.~Fusaoka for helpful discussions. 
Especially, he thanks H.~Fusaoka for pointing out  
incompleteness of the scenario in the earlier version.
This work is supported by the Grant-in-Aid for
Scientific Research (C), JSPS, (No.21540266).




\vspace{4mm}
\begin{thebibliography}{99}
%
\bibitem{mnu-yukawaonPRD08} Y.~Koide, Phys.~Rev. {\bf D78} (2008) 093006 .
%
\bibitem{me-yukawaonPRD09} Y.~Koide, Phys.~Rev. {\bf D79} (2009) 033009.
%
%
\bibitem{Koide-O3-PLB08} Y.~Koide, Phys.~Lett. {\bf B665} (2008) 227.
%
%
%
%

%
\bibitem{CKM_SD}
L.~-L.~Chau and W.~-Y.~Keung, Phys.~Rev.~Lett. {\bf 53}, 1802 (1984);
H.~Fritzsch, Phys.~Rev. {\bf D32} (1985) 3058.
%
%
\bibitem{PDG08} C.~Amsler, {\it et al}., Particle Data Group, 
Phys.~Lett. {\bf B667} (2008) 1.
%
\bibitem{tribi} 
P.~F.~Harrison, D.~H.~Perkins and W.~G.~Scott,
 Phys.~Lett. {\bf B458} (1999)  79;
 Phys.~Lett. {\bf B530} (2002) 167;
Z.-z.~Xing, Phys.~Lett. {\bf B533} (2002) 85;
P.~F.~Harrison and W.~G.~Scott,  Phys.~Lett. {\bf B535} (2002) 163;
Phys.~Lett. {\bf B557} (2003)76;
E.~Ma, Phys.~Rev.~Lett. {\bf 90} (2003) 221802;
C.~I.~Low and R.~R.~Volkas, Phys.~Rev. {\bf D68} (2003) 033007.
%
\bibitem{democ-seesaw} Y.~Koide and H.~Fusaoka, Z.~Phys. {\bf C71}
 (1996) 459; Prog.~Theor.~Phys. {\bf 97}, (1997) 459.
%
%
\bibitem{Froggatt} C.~D.~Froggatt and H.~B.~Nielsen, 
Nucl.~Phys. {\bf B 147} (1979) 277.
%
\bibitem{e-Yukawaon09} Y.~Koide, arXiv:0906.3370 [hrp-ph].
%
\bibitem{noYnu} Y.~Koide, arXiv:0812.3203 [hep-ph], a talk at  
``Particle Physics, Astrophysics and Quantum Field Theory: 75 Years 
since Solvay" (PAQFT08), 27-29 Nov. 2008, Nanyang Executive Centre, Singapore,
To appear in the Conference Proceedings (Intl.J.Mod.Phys.A).
%
%
\bibitem{q-mass} Z.-z.~Xing, H.~Zhang and S.~Zhou, 
Phys.~Rev. {\bf D77} (2008) 113016.
Also, see H.~Fusaoka and Y.~Koide, Phys. Rev. 
{\bf D57} (1998) 3986.
%
\bibitem{Koidemass} Y.~Koide, Lett.~Nuovo Cimento {\bf 34} (1982) 201;
Phys.~Lett. {\bf B120} (1983) 161;
Phys.~Rev. {\bf D28} (1983) 252.
%
%
\bibitem{MINOS} D.~G.~Michael {\it et al.}, MINOS collaboration,
Phys.~Rev.~Lett. {\bf 97} (2006) 191801;
J.~Hosaka, {\it et al.}, Super-Kamiokande collaboration, Phys.~Rev. 
{\bf D74} (2006) 032002.
%
\bibitem{SNO08} B.~Aharmim, {\it et al.}, SNO collaboration,
Phys.~Rev.~Lett. {\bf 101} (2008) 111301.
Also, see S.~Abe, {\it et al.}, KamLAND collaboration,
Phys.~Rev.~Lett. {\bf 100} (2008) 221803.
%
\end{thebibliography}
\end{document}